\documentclass{ws-ijmpa}
\usepackage[dvips]{color}

\usepackage{color}
\usepackage{graphicx}

\newcommand{\eps}{\varepsilon}

\begin{document}

\title{Tensorial NSI and Unparticle physics in neutrino
  scattering} 

\author{J. Barranco$^{1,2}$}
\address{$^1$Instituto de Astronom\'{\i}a, Universidad Nacional
  Aut\'onoma de M\'exico,\\ M\'exico, DF 04510, Mexico\\$^2$Divisi\'on de Ciencias e Ingenier\'ias,  Universidad de Guanajuato,\\
Campus Leon, C.P. 37150, Le\'on, Guanajuato, M\'exico\footnote{jbarranc@fisica.ugto.mx}}
\author{A. Bola\~nos$^3$}
\address{$^3$Facultad de Ciencias F\'\i sico Matem\' aticas, Benem\' erita Universidad Aut\' onoma de Puebla,
\\ Apartado Postal 1152, Puebla, Puebla, M\'exico. \footnote{azucena@fcfm.buap.mx}}
\author{E. A. Garc\'es$^4$}
\address{$^4$Departamento de F\'{\i}sica, Centro de Investigaci{\'o}n y de Estudios Avanzados del IPN, \\Apdo. Postal 14-740 07000
M\'exico, D F, Mexico\footnote{egarces@fis.cinvestav.mx}}

\author{O. G. Miranda$^4$}
\address{$^4$Departamento de F\'{\i}sica, Centro de Investigaci{\'o}n y de Estudios Avanzados del IPN, \\Apdo. Postal 14-740 07000
M\'exico, D F, Mexico\footnote{omr@fis.cinvestav.mx}}
\author{T. I. Rashba$^5$}
\address{$^5$IZMIRAN,
Institute of Terrestrial Magnetism, Ionosphere and Radio Wave Propagation of the Russian Academy of Sciences,
142190, Troitsk, Moscow region, Russia\footnote{timur@mppmu.mpg.de}}

\maketitle
\begin{abstract}
We have analyzed the electron anti-neutrino scattering off electrons
and the electron anti-neutrino-nuclei coherent scattering in order to
obtain constraints on tensorial couplings.  We have studied the
formalism of non$-$standard interactions (NSI), as well as the case of
Unparticle physics.  For our analysis we have focused on the recent
TEXONO collaboration results and we have obtained current constraints
to possible electron anti-neutrino-electron tensorial couplings in
both new physics formalisms.  The possibility of measuring for the
first time electron anti-neutrino-nucleus coherent scattering and its
potential to further constrain electron anti-neutrino$-$quark
tensorial couplings is also discussed.
\end{abstract}
\ccode{PACS numbers:13.15.+g,12.90.+b,14.60.St}

\section{Introduction}

Searching for new physics has been highly motivated by neutrino
oscillation physics which implies non-zero neutrino
mass~\cite{Schwetz:2011qt}.  Neutrino oscillations are the first clue
of physics beyond the standard model and this fact has motivated a new
generation of neutrino experiments as well as the development of
phenomenological and theoretical research on models beyond the
Standard Model~\cite{Nunokawa:2007qh,Mohapatra:2006gs} . The future
generation of neutrino experiments put the field as very promising in
the search for new physics.

A very useful tool for the phenomenological study of these types of
new physics is the formalism of non standard interactions (NSI). This
formalism can parametrize a wide range of well motivated models of
neutrino mass and, at the same time, gives model
independent constraints.

In the NSI formalism, operators with (V$\pm$A) Lorentz structure have
been widely studied in the
literature~\cite{Barger:1991ae,Bergmann:2000gp,Huber:2001zw,berezhiani,davidson},
giving stringent limits in both neutrino-electron and
anti-neutrino-electron interactions when solar and reactor data are
analyzed~\cite{Barranco:2005ps,Barranco:2007ej,Bolanos:2008km,Forero:2011zz,Escrihuela:2011cf,Kopp:2010qt,Ribeiro:2007jq,Huber:2006wb}.

However, a similar analysis based on an effective approach of
tensorial NSI has not been done in the neutrino sector.
To the best of our knowledge, constraints on tensorial NSI based on
this phenomenological approach have been only derived using
accelerator neutrinos at LAMPF~\cite{Allen:1992qe} and stellar energy
loss arguments~\cite{Dicus:1976ra,Sutherland:1975dr}. In the present
work we use the most recent measurement on the $\bar \nu_e$-electron
scattering to constrain the tensorial NSI coupling constant.

The study of neutrino-electron scattering has been of interest for
elementary particle physics since its first measurement at the
GARGAMELLE bubble chamber~\cite{Hasert:1974}. Recently, neutrino
detectors using reactor  anti-neutrinos in a short baseline have taken
profit of the intense and pure source of $\bar \nu_e$ as a probe of
physics beyond the Standard
Model~\cite{Daraktchieva:2005kn,Wong:2006nx,Deniz:2009mu}.

Moreover, $\bar \nu_e$-electron scattering has been recently measured with
improved accuracy by using reactor anti-neutrinos. In particular the
TEXONO collaboration has reported recent
results~\cite{Deniz:2009mu,Deniz:2010mp} 
  using $187$~kg of
  crystal scintillator detector ($CSI(Tl)$). In their results they
  showed,contrary to previous experiments, a binned sample for low
  energy antineutrino electron scattering. It is natural to expect a
  better accuracy in searching for new physics by using this data
  sample. 
 In the present work we have
used these results to study tensorial interactions in the framework of
NSI.

On the other hand, a very different type of new physics that also can
give rise to tensorial interactions is the recently proposed
unparticle physics. In this extension, particles couple to a hidden
conformal sector~\cite{Georgi:2007ek,Georgi:2007si} which could be
probed in future experiments.

Unparticle physics signatures could be directly produced in
accelerator experiments and therefore tested by searching for
signatures of missing energy in the detectors.  There is another way
to study their effects, which is through low energy processes mediated
by the unparticle stuff. The latter could, for example, modify
neutrino elastic scattering phenomenology due to effects of virtual
unparticle exchange between fermionic currents.

Previous analysis of $\bar \nu_e$-electron elastic scattering 
have been performed for scalar and vectorial unparticle
propagators~\cite{Zhou:2007zq,Balantekin:2007eg,Barranco:2009px} in
order to obtain constraints on relevant unparticle parameters.  Here
we introduce the analysis of tensorial unparticle propagators in
$\bar \nu_e$ elastic scattering off electrons and in $\bar \nu_e$-nuclei
coherent scattering.

We will see in this work that it is possible to obtain strong
constraints to both NSI and unparticle parameters using
$\bar \nu_e$-electron scattering results. These constraints are valid for
a non conformal invariant unparticle theory where the dimension of the
operator is not bounded from unitarity
constraints~\cite{GonzalezGarcia:2008wk,Fortin:2011sz}.  Furthermore,
the unparticle tensor propagator used in our analysis is antisymmetric
as the one reported in Ref.~\refcite{Hur:2007cr}.

Our paper is organized as follows: In Section \ref{nsi} we introduce
the NSI formalism and present the corresponding $\bar \nu_e$-electron and
the $\bar \nu_e$-nucleus coherent scattering cross sections, while in
Section \ref{unparticle} we present the analog discussion for the
unparticle case. The description of our analysis for the TEXONO case
is shown in Section \ref{texono},  and for the $\bar \nu_e$-nucleus coherent
  scattering is shown in Section \ref{coherent}.  Finally, our
conclusions are given in Section \ref{conclusions}.

\section{Tensorial NSI in neutrino scattering}
\label{nsi}

It is a common characteristic of many extensions of the Standard Model (SM)
to introduce new interactions that can be parametrized with the help
of an effective Lagrangian. The most studied case regards with $(V-A)$
extensions of the SM. It is well known that in this case
the effective four fermion Lagrangian takes the
form:~\cite{Barger:1991ae,Bergmann:2000gp,Huber:2001zw,berezhiani,davidson}

\begin{equation}
-{\cal L}^{eff}_{\rm V-A} =
\eps_{\alpha \beta}^{fP}{2\sqrt2 G_F} (\bar{\nu}_\alpha \gamma_\rho L 
\nu_\beta)
( \bar {f} \gamma^\rho P f ), \label{lagrangian_nsi}
\end{equation}
where $f$ is a first generation SM fermion: $e,u$ or $d$, and $P=L$ or
$R$, are the chiral projectors and $\eps_{\alpha \beta}^{fP}$ parametrize
the strength of the NSI with $\alpha$ and $\beta$ the initial and
final flavor states.  There is plenty of literature studying
either current constraints on the neutrino NSI with
electrons~\cite{Barranco:2005ps,Barranco:2007ej,Bolanos:2008km,Forero:2011zz} 
and quarks~\cite{Escrihuela:2011cf,Kopp:2010qt} as well as future
perspectives for long baseline neutrino
experiments~\cite{Ribeiro:2007jq,Huber:2006wb} as well as in reactor
low energy neutrino experiments~\cite{Barranco:2005yy}.

In particular for the $\nu_e$-electron scattering, which has the
advantage of being a purely leptonic process and therefore is free
from QCD uncertainties, the $\nu_e$-electron cross section for the
NSI case is given by 
\begin{eqnarray} 
\frac{d\sigma} {dT} &=& {2 G_F^2 m_e \over \pi} [ (\tilde g_L^2+\sum_{\alpha \neq e}
|\epsilon_{\alpha e}^{e L}|^2) + (\tilde g_R^2+\sum_{\alpha \neq e} |\epsilon_{\alpha e}^{e R}|^2)\left(1-{T \over E_{\nu}}\right)^2
 \nonumber\\ &-& (\tilde g_L \tilde g_R + 
\sum_{\alpha \neq e}|\epsilon_{\alpha e}^{e L}||
\epsilon_{\alpha e}^{e R}|)m_e {T \over E_{\nu}^2}] ,
\label{cross-section}
\end{eqnarray} 
Here, $E_\nu$ is the incident neutrino energy, $T$ is the
  electron recoil energy, $m_e$ is the electron mass, $\tilde
g_L=g_L+\epsilon_{e e}^{e L}$, $\tilde
g_R=g_R+\epsilon_{e e}^{e R}$   
and $G_F$ the Fermi constant.
The SM coupling constants are given by $g_{L}=\frac{1}{2}+\sin^{2}\theta_{W}$,
$g_{R}=\sin^{2}\theta_{W}$. 
For $\bar \nu_e$-electron the cross section is 
obtained by interchanging $\tilde g_{L(R)} \to \tilde g_{R(L)}$ in Eq. 
\ref{cross-section}.
It can be noticed that, in the 
absence of NSI, this expression takes the form of the SM cross section. 

On the other hand, limited attention has been given to possible tensor
interactions. This may be motivated by the $(V-A)$ structure of the
SM, however, with neutrino physics entering into a
precision era, it could be a good moment to study this kind of
interactions with more detail.
In previous studies, the tensorial fermion currents of the form
$\bar{f}\sigma_{\mu\nu} f $ have been
studied~\cite{Kingsley:1974kq,Cho:1976um}. In particular,
$\nu_e$-electron scattering received some
attention~\cite{Kayser:1979mj} and there have been constraints
reported in the literature~\cite{Allen:1992qe}.

In the case of $\bar \nu_e$ scattering off
electrons, the tensorial contribution to the amplitude is given
by~\cite{Kayser:1979mj}
\begin{equation}
\mid M\mid^{2}= \sum_{\beta=e,\mu,\tau}{\eps_{e \beta}^{eT}}^{2}\frac{G_{F}^{2}}{2}128 m_e^{2}(4E_{\nu}^{2}+T^{2}-(4E_{\nu}+m_e)T)
\end{equation}
and therefore the differential NSI $\bar \nu_e$-electron cross section
takes the form:
\begin{equation}
\frac{d\sigma^{NSI}_{T}}{dT}= \frac{\mid M\mid^{2}}{64\pi m_e E_{\nu}^{2}}
={\sum_{\beta=e,\mu\tau}\eps_{e \beta}^{eT}}^{2}\frac{4G_{F}^{2}m_e}{\pi}
\Big[ \Big(1-\frac{T}{2E_{\nu}}\Big)^{2}-\frac{m_eT}{4E_{\nu}^{2}}\Big] .
\label{cs_nsi}
\end{equation}
In this case, $\eps_{e \beta}^{eT}$ parametrizes the strength of 
the tensorial NSI coupling on electrons. 
Note that there is no interference term between the tensorial non
standard amplitude and the SM amplitude. Neutrino
flavor changing processes have the same contribution to the total
cross section as the conserving neutrino flavor process. Hence, the
limits that are derived for $\eps_{e e}^{eT}$ are the same as the
ones obtained for flavor changing coupling constants $\eps_{e\beta}$
$\beta \ne e$  when 
only one parameter is allowed to vary at a time.  
For this reason, from now on we will
denote the tensorial coupling $\eps_{e \beta}^{eT}$ as $g^{T_e} $. It is
also worth to be noticed that the tensor interaction could imply a
change in chirality for the incoming neutrino. Given the fact that
neutrinos are massive it is natural to consider such a possibility.  

\begin{figure}
\begin{center}
\includegraphics[angle=0,width=0.5\textwidth]{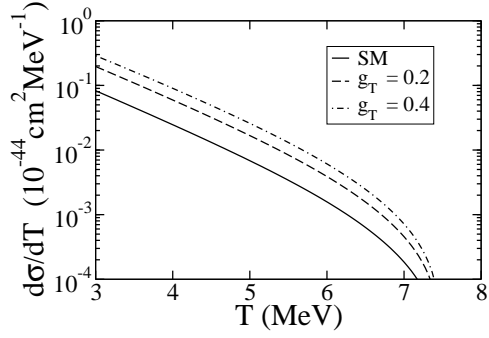}
\end{center}
\caption{Comparison between Standard Model and
  tensor NSI differential cross section.}
\label{fig:cs}
\end{figure}

In order to have an idea about the effect of the tensorial  NSI interaction, 
we plot the differential cross section versus the
energy and compare it with the SM prediction,
figure (\ref{fig:cs}), for some values of $g^{T_e} $. 

\subsection{Neutrino-nucleus coherent scattering}
The coherent neutrino-nucleus~\cite{Freedman:1973yd} and
neutrino-atom~\cite{Gaponov:1977gr,Sehgal:1986gn} scattering have been
recognized for many years as interesting processes to probe the SM.
The coherent scattering takes place when momentum transfer, $q$, is
small compared with $R^{-1}$, the inverse nucleus (or atom) size, i.e. $qR
< 1$. For most nuclei, this condition is fulfilled for neutrino
energies below 150 MeV. Therefore, the condition for full coherence in
the neutrino-nucleus scattering is well satisfied for reactor
anti-neutrinos and also for solar, supernovae, and artificial neutrino
sources.

The TEXONO collaboration has a program towards the detection of the
coherent $\bar \nu_e$-nucleus scattering~\cite{TEXONO0402}. Such a
detection will be helpful to improve the limits on the neutrino
magnetic moment and other types of new physics such as NSI and
unparticle tensorial interactions and might even set better
constraints than those coming from future neutrino factory
experiments~\cite{Barranco:2005yy}.

Within the SM,  neglecting radiative corrections, 
the cross section for $\bar \nu_e$-nucleus coherent scattering is
\begin{eqnarray}
\frac{d\sigma}{dT}=\frac{G_F^2 M}{2\pi} \left\{
(G_V-G_A)^2+\left(G_V+G_A\right)^2\left(1-\frac{T}{E_\nu}\right)^2 - 
\left(G_V^2-G_A^2\right) \frac{MT}{E_\nu^2} \right\},\label{diff:cross:sect}
\end{eqnarray}
where $M$ is the mass of the nucleus, $T$ is the recoil nucleus energy, $E_\nu$ is the incident anti-neutrino energy and the axial and vector couplings are
\begin{eqnarray}
\label{GV}
G_V&=& 
\left[g_V^p Z+g_V^nN\right]
F_{nucl}^V(q^2)\,,\\
G_A&=& 
\left[g_A^p \left(Z_+-Z_-\right)+g_A^n\left(N_+-N_-\right)\right]
F_{nucl}^A(q^2)\,.
\label{GA}
\end{eqnarray}
$Z$ and $N$ represent the number of protons and neutrons in the
nucleus, while $Z_{\pm}$ ($N_\pm$) stands for the number of protons
(neutrons) with spin up and spin down respectively.
The vector and axial nuclear form factors, $F_{nucl}^V(q^2)$ and
$F_{nucl}^A(q^2)$, are usually assumed to be equal and of order of
unity in the limit of small energies, $q^2\ll M^2$.
The SM neutral current vector couplings of neutrinos with
protons, $g_V^p$, and with neutrons, $g_V^n$, are 
\begin{eqnarray}
&&g_V^p=\rho_{\nu N}^{NC}\left(
\frac12-2\hat\kappa_{\nu N}\hat s_Z^2
\right)+
2\lambda^{uL}+2\lambda^{uR}+\lambda^{dL}+\lambda^{dR},\nonumber\\
&&g_V^n=-\frac12\rho_{\nu N}^{NC}+
\lambda^{uL}+\lambda^{uR}+2\lambda^{dL}+2\lambda^{dR}\,.
\label{vcouplings}
\end{eqnarray}

Here $\hat s_Z^2=\sin^2\theta_W=0.23120$, $\rho_{\nu N}^{NC}=1.0086$,
$\hat\kappa_{\nu N}=0.9978$, $\lambda^{uL}=-0.0031$,
$\lambda^{dL}=-0.0025$ and $\lambda^{dR}=2\lambda^{uR}=7.5\times10^{-5}$
are the radiative corrections given by the 
PDG~\cite{Yao:2006px}. 
The axial contribution can be neglected as can be seen from Eq. (\ref{diff:cross:sect}) since 
the ratio of axial to vector contribution is expected to be of the order $1/A$,  where $A$ is the atomic number. 
The spin-zero cross section of $\bar \nu_e$ scattering off nuclei in the 
low energy limit, $T \ll E_{\nu}$ is 
\begin{equation}
\frac{d\sigma}{dT}=\frac{G_F^2 M}{\pi}
\left(1-\frac{M T}{2E_\nu^2}\right)\left[
Z (g_V^p)+N (g_V^{n})\right]^2.\label{CS}
\end{equation}

Now we can compute the $\bar \nu_e$-nucleus coherent dispersion with a
tensorial NSI coupling.  Analogously to the previous lines, 
and incorporating the tensorial NSI term,
we calculate the tensor NSI coherent neutrino-nucleus cross section
as:
\begin{equation}\label{nsi_coh}
\frac{d\sigma_{T}^{NSI}}{dT}=
 \frac{4G_F^2 M}{\pi}
\left[g^{T_u}(2Z+N)+g^{T_d}(Z+2N)\right]^2
\left[\left(1-\frac{T}{2 E_\nu}\right)^2-\frac{M T}{4 E_\nu^2}\right]\,,
\end{equation}
where $m_A$ is the mass of the nucleus and $g^{T_u}$, $g^{T_u}$ the
  tensor couplings for u-type or d-type quark, respectively.

\section{Neutrino-electron scattering mediated by tensorial 
unparticle interactions}
\label{unparticle}

At energies above certain $\Lambda$, a hidden sector operator ${\cal
  O}_{UV}$ of dimension $d_{UV}$ could couple to the SM operators
${\cal O}_{SM}$ of dimension $d_{SM}$ via the exchange of heavy
particles of mass $M$
\begin{equation}
{\cal L}_{UV}=\frac{{\cal O}_{UV}{\cal O}_{SM}}{M^{d_{UV}+d_{SM}-4}}\,.
\end{equation}
The hidden sector becomes scale invariant at $\Lambda$ and then the
interactions become of the form
\begin{equation}
{\cal L}_{\cal U}=
C_{\cal O_U}\frac{\Lambda^{d_{UV}-d}}{M^{d_{UV}+d_{SM}-4}}\,{\cal O}_{\cal
  U}\,{\cal O}_{SM}\,,
\end{equation}
where ${\cal O}_{\cal U}$ is the unparticle operator of scaling
dimension $d$ in the low energy limit and $C_{\cal O_U}$ is a
dimensionless coupling constant.  Therefore the unparticle sector can
appear at low energies in the form of new massless fields coupled very
weakly to the SM particles.

Scalar and vectorial unparticle interactions have been studied
previously in the context of $\bar \nu_e$-electron
scattering~\cite{Deniz:2010mp,Balantekin:2007eg,Barranco:2009px}. In
what follows we will concentrate on the case of tensorial
interactions.

Effective interactions for the tensor unparticle interactions in the low 
energy regime have been studied in the past~\cite{Hur:2007cr,Cheung:2007ap}. 

In this work we will use the antisymmetric tensor operator of the 
form~\cite{Hur:2007cr} 
\begin{equation}
[\mathcal{A}_\mathcal{F}(P^{2})]_{\mu\nu,\rho\sigma}=\frac{\mathcal{A}_{d}}{2\sin{(d}\pi)} 
(-P^{2})^{d-2}T_{\mu\nu\rho\sigma}(P),
\end{equation}
where
\begin{equation}\label{Ad}
\mathcal{A}_{d}= \frac{16\pi^{5/2}}{(2\pi)^{2d}}
\frac{\Gamma(d+1/2)}{\Gamma(d-1)\Gamma(2d)} .
\end{equation}
The tensor $T_{\mu\nu\rho\sigma}$ is split into a 'magnetic' and an
'electric part'.  The 'magnetic part' (m) is defined as:
\begin{equation}
\label{Tmagn}
T_{\mu\nu\rho\sigma}^{(m)}=\frac12\left\{
\pi_{\mu\rho} \pi_{\nu\sigma}-
\pi_{\mu\sigma} \pi_{\nu\rho}
\right\},
\end{equation}
and the 'electric part' is given by:
\begin{equation}
\label{Telec}
T_{\mu\nu\rho\sigma}^{(e)}=\frac12\left\{
\pi_{\mu\rho} w_{\nu\sigma}-
\pi_{\mu\sigma} w_{\nu\rho}- 
\pi_{\nu\rho} w_{\mu\sigma}+
\pi_{\nu\sigma} w_{\mu\rho} 
\right\},
\end{equation}
where 
\begin{equation}
w_{\mu\nu}=\frac{(k-k')_\mu(k-k')_\nu}{(k-k')^2}\,.
\end{equation}

The neutrino matrix for the neutrino (of any flavor)-fermion interaction, mediated
by a tensor unparticle is 
\begin{equation}\label{tensor_amplitude}
\mathcal{M}=\frac{\lambda_\nu^{\alpha \beta} \tilde \lambda_f}{2\sin(d\pi)\Lambda_U^{2d-2}}A_d
[\bar \nu_\beta(k')\sigma^{\mu\nu}\nu_\alpha(k)](-(k-k')^2)^{d-2}T_{\mu\nu\rho\sigma}
[\bar f(p')\sigma^{\rho\sigma}f(p)]\,,
\end{equation} 
and $T_{\mu\nu\rho\sigma}$ is either the 'magnetic' $T^{(m)}$ or
the 'electric' part $T^{(e)}$.
In what follows we will use the definition for the neutrino and 
fermion couplings as 
$\lambda_f=\sqrt{\lambda_{\nu}^{\alpha\beta}\tilde \lambda_{f}}$, with 
\begin{equation}
\lambda_{\nu}^{\alpha\beta} = C^{\alpha\beta}_{\cal O_U \nu}
\frac{\Lambda^{d_{UV}-d}}{M^{d_{UV}+d_{SM}-4}} \,;\quad
\tilde \lambda_{f} = C_{{\cal O_U} f }
\frac{\Lambda^{d_{UV}-d}}{M^{d_{UV}+d_{SM}-4}} \,,
\end{equation}
and we will fix the scale $\Lambda = 1$~TeV.  With this information, it is
possible to obtain the differential cross section for $\bar
\nu_e$-electron scattering for the tensorial unparticle
interactions. We concentrate in the flavor conserving case of the
$\bar \nu_e$-electron interaction,
i.e. $\lambda_e=\sqrt{\lambda_{\nu}^{e e}\tilde \lambda_{e}}$.

The 'electric part' contributes with the differential cross section

\begin{equation}
\frac{d\sigma_T}{dT}=
\frac{f(d)^2}{\pi \Lambda_U^{4d-4}}
2^{2d-3}m_e^{2d-3}T^{2d-4}
\Big[\Big(1-\frac{T}{2E_{\nu}}\Big)^{2}-\frac{m_eT}{2E_{\nu}^{2}}\Big],
\end{equation}
and the 'magnetic part' contribution is given by
\begin{equation}
\frac{d\sigma_T}{dT}=
\frac{f(d)^2}{\pi \Lambda_U^{4d-4}}
2^{2d-2}m_e^{2d-3}T^{2d-4}
\Big(1-\frac{T}{2E_{\nu}}\Big)^{2},
\label{magnetic}
\end{equation}
where we have defined
\begin{equation}
f(d)=\frac{\lambda_e}{2\sin(d\pi)}A_d .
\label{eq:f(d)}
\end{equation}
Notice that in this case an integer dimension for $d$ leads to a
  singularity in the value of $f(d)$.

In order to obtain the total cross section, both expressions, the
'magnetic' and 'electric' contributions should be added to the
SM prediction

\begin{equation}
\frac{d\sigma(\bar{\nu_e})}{dT}= \frac{2G_{F}^{2}m_{e}}{\pi}[g_{R}^{2}+
g_{L}^{2}(1-\frac{T}{E_{\nu}})^{2}-g_{R}g_{L}\frac{m_{e}T}{E_{\nu}^2}] .
\label{electric}
\end{equation}

\begin{figure}
\begin{center}
\includegraphics[angle=0,width=0.5\textwidth]{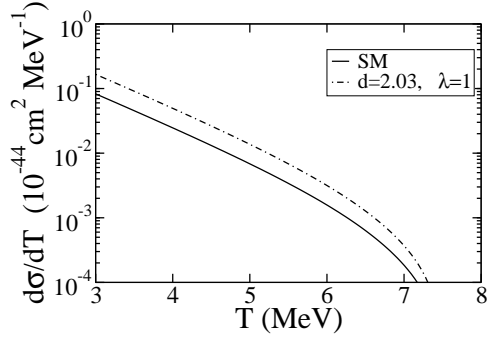}
\end{center}
\caption{Comparison between Standard Model and tensor unparticle
  differential cross section for $\Lambda=1$~TeV. }
\label{fig:cs_unparticle}
\end{figure}

We show in figure (\ref{fig:cs_unparticle}) the comparison between the
SM prediction for the differential cross section and the
unparticle case, we can see that the
expectations to obtain a good constraint from this process at low
energies are encouraging.

If the theory is not only scale invariant but also conformal invariant
then unitarity constraints apply to the the dimension d of the
unparticle operator~\cite{Grinstein:2008qk}.  We will relax this
constraint in our phenomenological analysis as has been done in other
phenomenological and theoretical
works~\cite{GonzalezGarcia:2008wk,Fortin:2011sz}.

\subsection{$\bar \nu_e$-nucleus coherent scattering for tensorial unparticle}

Now we can write the tensor unparticle part of the $\bar \nu_e$-nucleus coherent
scattering cross section. 
We get the following expressions for the 'magnetic part'
\begin{eqnarray}\label{unparticle_coh_mag}
\frac{d\sigma_{U_T}^{(m)}}{dT}&=&
\frac{1}{\pi \Lambda_u^{4d-4}}
\left[g_{u}(d)(2Z+N)+g_{d}(d)(Z+2N)\right]^2\times\nonumber\\
&&
2^{2d-2} m_A^{2d-3} T^{2d-4}
\left(1-\frac{T}{2E_\nu}\right)^2\,,
\end{eqnarray}
while for the 'electric part' we get 
\begin{eqnarray}\label{unparticle_coh_elec}
\frac{d\sigma_{U_T}^{(e)}}{dT}&=&
\frac{1}{\pi \Lambda_u^{4d-4}}
\left[g_{u}(d)(2Z+N)+g_{d}(d)(Z+2N)\right]^2\times\nonumber\\
&&
2^{2d-3} m_A^{2d-3} T^{2d-4}
\left(\left(1-\frac{T}{2E_\nu}\right)^2-\frac{m_A T}{2E_\nu^2}\right)\,.
\end{eqnarray}
In the last expressions we have defined the new coupling constants
\begin{equation}
g_{u,d}(d)=\frac{\lambda_{i\nu}^{e e}\tilde \lambda_{u,d}}{2\sin(d\pi)}~A_d=
\frac{\lambda_{u,d}^2}{2\sin(d\pi)}A_d\,,
\end{equation}
where we have used the same definition of $A_d$ as defined for the 
$\bar \nu_e$-electron scattering.

\section{Limits on tensor interactions from the TEXONO experiment}
\label{texono}

Among the most recent reactor neutrino experiments, the TEXONO
collaboration has published results on the cross section for the
$\bar \nu_e$-electron scattering~\cite{Chang:2006ug,Deniz:2010mp} using
the Kuo Sheng 2.9 GW reactor as an anti-neutrino source that provides an 
average flux of $6.4\times 10^{12}cm^{-2} s^{-1}$. Even though the
collaboration has made use of three different detectors, we will focus
on the CsI(Tl) detector data in order to obtain constraints for the
tensor interactions both for the NSI and unparticle cases.

In order to obtain a constraint on the tensorial parameters we have
computed the expected number of events for the TEXONO detector in the
case of a NSI or unparticle interaction given by (\ref{cs_nsi})
and (\ref{magnetic},\ref{electric}) respectively and compute the integral 
\begin{equation}
N_i = K \int^{T_{i+1}}_{T_i} \int_{E_\nu} \frac{d\sigma}{dT} \frac{d\phi(\bar{\nu}_e)}{dE_\nu} dE_\nu dT ,
\end{equation}
where the factor $K$ accounts for the time exposure and the number of
electron targets, $d\sigma/dT$ is the cross section for the NSI or the
unparticle interaction and $d\phi(\bar{\nu}_e)/dE_\nu$ is the neutrino
spectrum which we have parametrized as the exponential of a polynomial
order five as has been recently discussed in the
literature~\cite{Mueller:2011nm}.  We have also considered the
relative abundances of each radioactive isotope in the nuclear
reactor, $^{235}$U($98\%$), $^{238}$U($1.5\%$), and
$^{239}$Pu($0.4\%$). The electron recoil energy is divided into ten
bins, $T_i$, running from $3$ to $8$~MeV. The detector is
  located at a distance of 28 m from the reactor.  For $\bar \nu_e$
  energies around $1~$MeV, the estimated oscillation length into an 
  active neutrino is of the
  order of $10~$km, hence in the calculation of the expected number
  of events we do not take into account neutrino oscillation effects.

Once we have computed the theoretical expected events per bin we can compute 
the $\chi^2$ function 

\begin{equation}
\chi^2 = \sum_{i=1} \left[  \frac{N_{expt}(i)-\left[ N_{NSI,U}(i)\right] }{\Delta_{stat}(i)} \right]^2,
\end{equation}
\begin{figure}
\begin{center}
  \includegraphics[angle=0,width=0.5\textwidth]{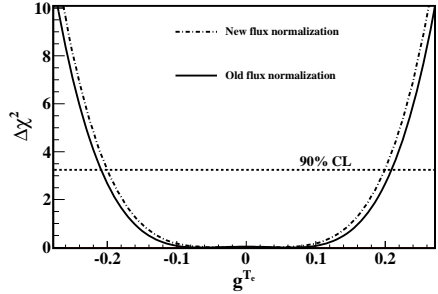}
\end{center}
\caption{$\Delta \chi^2$ for the tensorial 
NSI coupling  $g^{T_e} $ in the TEXONO experiment.}\label{fig:chi}
\end{figure}
where $N_{NSI,U}(i)$, is the calculated event rate in the $i$th energy
data bin for the Tensorial NSI or the unparticle cases, $N_{expt}(i)$ is
the observed event rate for the corresponding energy bin, and
$\Delta_{stat}(i)$ is the statistical uncertainty of the associated measurement.

\begin{figure}
\begin{center}
\includegraphics[angle=0,width=0.5\textwidth]{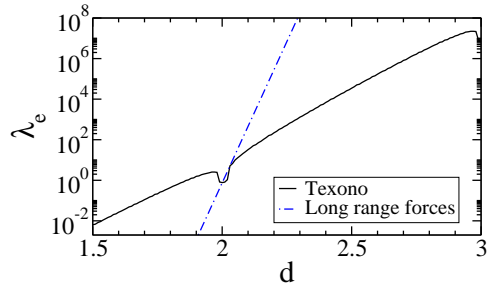}
\end{center}
\caption{limits at 90 \% CL for the tensorial unparticle parameters $d$ and $\lambda_e$ from our 
analysis of the recent TEXONO data. }
\label{fig:chi_unparticle}
\end{figure}

The results of our analysis are shown in figures~(\ref{fig:chi}) and
(\ref{fig:chi_unparticle}). We can see that for the NSI case the
constraint on the tensorial coupling gives the bound $g^{T_e} \leq
0.20$ at 90 \% C. L., which is much better than the previously
reported constraint by the LAMPF collaboration~\cite{Allen:1992qe}. We
obtained the value of $\chi^2_{min} = 5.43$ for $9$~d.o.f.  We have
verified explicitly that previous reactor experiments give weaker
constraints than those obtained here using the TEXONO data (see Table
\ref{table:nsi}). Recently, a recalculation of the reactor
  anti-neutrino fluxes has been done~\cite{Mention:2011rk}, leading to
  a deficit in the observed rates of reactor neutrino experiments; we
  have also performed an analysis taken into account this
  revaluation. The result is shown in figure~(\ref{fig:chi}). We can
  see that there is some impact in the constraints, although we prefer
  to be conservative and quote the more relaxed bounds that are still
  better than previous reported constraints.

For the unparticle case we have shown the $90$ \% CL region for the
parameter $\lambda_e$ and $d$. In this case we obtained $\chi^2_{min} =
5.16$ for $8$~d.o.f. We have also extracted the constraints from the
solar neutrino analysis reported in
Ref.~\refcite{GonzalezGarcia:2008wk} (following analogous
assumptions to those discussed in Ref.~\refcite{Barranco:2009px}) and also
plotted the result in figure~(\ref{fig:chi_unparticle}) in
order to show the interplay between different analysis; as can be seen,
the results from our analysis are more restrictive for values of
$d>2.03$.
We also can note, as expected from Eq.~(\ref{eq:f(d)}), that 
there is a singularity for integer values of $d$, for example, in the case 
of $d=2$.

\section{Tensorial neutrino-quark constraints with $\bar \nu_e$-nucleus coherent scattering }
\label{coherent}
In this section we study the sensitivity to tensorial NSI and
unparticle couplings coming from the coherent $\bar \nu_e$-nucleus
scattering. In order to apply our analysis to a concrete case, we will
concentrate our discussion on the germanium TEXONO
proposal~\cite{TEXONO0402}. The detector would be located at the
Kuo-Sheng Nuclear Power Station at a distance of 28 m from the reactor
core. We assume a typical neutrino flux of
$10^{13}$~s$^{-1}$~cm$^{-2}$.  Since the experiment is not running yet
and therefore we don't know the precise fuel composition, we use
  for this case the main component of the
spectrum~\cite{Schreckenbach:1985ep} coming from $^{235}$U. For
  energies below $2$~MeV there are only theoretical calculations for
  the anti-neutrino spectrum that we take from
  Ref.~\refcite{Kopeikin:1997ve}. We can estimate the expected total
  number of events in the detector in an analogous way as for the
  previous section
\begin{equation}
N_{\rm{events}}= K 
\int\limits_{E_{min}}^{E_{max}}dE_\nu
\int\limits_{T_{th}}^{T_{max}(E_\nu)}dT
\lambda(E_\nu)\frac{d\sigma}{dT}(E_\nu,T)\,,
\label{Nevents}
\end{equation}
where in this case $K=t\phi_0N_n$, with $t$ the data taking time
period, $\phi_0$ the total neutrino flux, and  $N_n$ the number of targets
in the detector, $\lambda(E_\nu)$ the normalized neutrino spectrum,
$E_{max}$ the maximum neutrino energy, $T_{th}$ the detector energy
threshold and the differential cross section refers to the coherent 
$\bar \nu_e$-nucleon interaction. Notice that in this case we are considering 
the total number of events without binning the sample and we are neglecting 
neutrino oscillation effects because the distance to the source is small 
compared with the typical oscillation length.

For the particular case of a minimum detector energy threshold of
$T_{th}=400$ eV, a 1 kg mass detector made of $^{76}$Ge and 1 yr of
data taking we found that the number of events is
$N_{events}^{SM}=4346$,  in good agreement with TEXONO
proposal~\cite{Wong:2011zzd}.

\begin{figure}
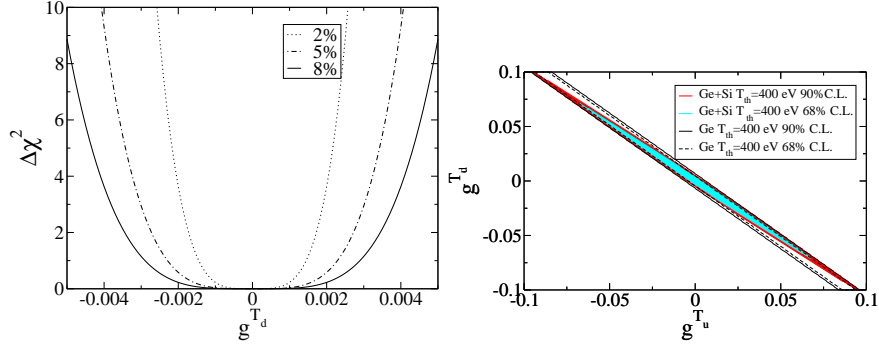

\begin{center}
\includegraphics[width=0.45\textwidth]{figura5a.eps}
\includegraphics[width=0.45\textwidth]{figura5b.eps}
\caption{Left panel: $\Delta \chi^2$ at $90\%$ CL for different total
  errors expected for TEXONO proposal.  We take a threshold
    energy $T_{th}=400~$eV and vary only the $g^{T_d}$ parameter.
    Right panel:
    The allowed regions of tensorial NSI parameters
    $g^{T_u}$ and $g^{T_d}$ are shown at $68\%$
    and $90\%$ CL for combined data from two detectors of $^{28}$Si
    and $^{76}$Ge (colored regions) and only for $^{76}$Ge (solid and
    dashed lines). 
    A $1$~kg mass and $1$~yr of data taking is assumed 
    for both detectors. Only statistical errors are taken into
    account.\label{fig:tensorial_coherent}}.
\end{center}
\end{figure}

\begin{figure}
\begin{center}
\includegraphics[width=0.5\textwidth]{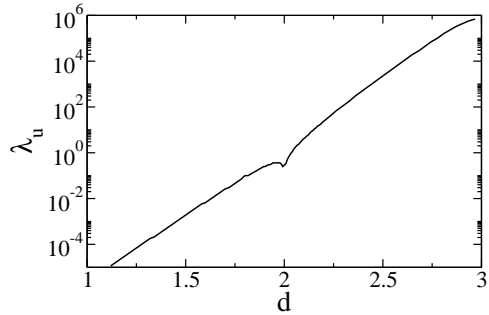}
\caption{Limits at $90\%$ CL for the unparticle case sensitivity on the parameters $d$ and $\lambda_u$ for the
neutrino-nucleus coherent scattering.}\label{unparticle_coh}
\end{center}
\end{figure}
We have estimated the sensitivity for the TEXONO
proposal to constrain unparticle parameters by means of a $\chi^2$
analysis
\begin{equation}
\chi^2=\left(\frac{N_{\rm{events}}^{\rm{SM}}-N_{\rm{events}}^{\rm NSI}}
 {\delta N_{\rm events}}\right)^2\,,
\end{equation}
where we have calculated $N_{\rm{events}}^{\rm NSI}$ by exchanging the
SM differential cross section in Eq.~(\ref{Nevents}) with the cross
section given in Eqs.~(\ref{unparticle_coh_mag}) and
(\ref{unparticle_coh_elec}), for the tensorial unparticle case and
with Eqs.~(\ref{nsi_coh}) for the tensorial NSI case respectively.
In the left panel of figure~(\ref{fig:tensorial_coherent}) we show the
$\Delta \chi^2$ function for the case when only one parameter
$g^{T_u}$ is varied at a time.  Furthermore, in right panel of the
same figure~(\ref{fig:tensorial_coherent}) we show in solid (dashed)
black lines the sensitivity for the tensorial non standard coupling at
$90\%$ ($68\%$) CL for the case of $1$ yr of data taking and a $1$ 
kg mass  $^{76}$Ge detector and varying both parameters
$g^{T_u}, g^{T_d}$.

As already discussed in Ref.~\refcite{Barranco:2005yy}, there is a degeneracy
in the parameters $g^{T_u},g^{T_d}$ as long as we use only one
material for the detector.  In order to break the degeneracy, another
material should be used.  We have proposed to use, in addition to 
the $^{76}$Ge detector, a $^{28}$Si detector to
break such degeneracy.  The expected sensitivity is shown in colored
lines in the same figure~(\ref{fig:tensorial_coherent}).

For the tensorial unparticle case, we  have performed a similar
analysis as done in Ref.~\refcite{Barranco:2009px} and we vary one parameter
at a time.  In figure~(\ref{unparticle_coh}) we show the sensitivity of
the coherent $\bar \nu_e-$nucleus  scattering for the tensorial unparticle
propagator for the case $\lambda_d=0$.

\section{Discussion and summary}
\label{conclusions}
Neutrino physics is entering into a precision era that could give
important guidance about new physics beyond the Standard Model.

In this article we have concentrated in the case of tensorial
couplings that could give a signal in reactor anti-neutrino
experiments. We have studied in particular the recent TEXONO results
on $\bar \nu_e$-electron scattering.  The tensorial interactions have
been studied both in the framework of NSI and for the unparticle case.
We have found new constraints that are stronger than previous
laboratory constraints.  These results can be summarized in Table
\ref{table:nsi} for the NSI case where we show the previous laboratory
result from the LAMPF experiment~\cite{Allen:1992qe}. Besides, we also
show the astrophysical estimates that come from stellar energy
loss~\cite{Dicus:1976ra,Sutherland:1975dr}.  For completeness, we
report the limits obtained by doing a $\chi^2$ analysis by using the
measurements of the cross section reported for the Irvine
experiment~\cite{Reines:1976pv} and MUNU~\cite{Daraktchieva:2003dr}.
\begin{table}[ph]
\tbl{Limits on the tensorial coupling  $g^{T_e} $, obtained
by using the data from previous experiments  
and from the TEXONO experiment analyzed in this work.
}
{\begin{tabular}{@{}cccc@{}} \toprule
experiment & Energy Range  (MeV) &Events &$g^{T_e} $  \\
\colrule
Stellar energy loss & $---$ & &$0.06 - 3.6$ \\
Irvine &$1.5-3.0$  & $381$ &$0.297$ $90$ \% CL\\ 
Irvine &$3.0-4.5$  & $77$&$0.360$ $90$ \% CL\\ 
LAMPF  &$10-50$ & $191$&$0.379$ $90$ \% CL\\ 
MUNU   &$0.7-2.0$ & $68$&$0.250$ $90$ \% CL\\
TEXONO &$3.0-8.0$ & $414$&$0.20$  $90$ \% CL\\ \botrule
\end{tabular}} \label{table:nsi}
\end{table}

\begin{table}[ph]
\tbl{Limits on the tensorial couplings $g^{T_d}$ and $g^{T_u}$, obtained
from a futuristic analysis of coherent neutrino-nucleus scattering
taking possible results from TEXONO as a case study.}
{\begin{tabular}{@{}ccc@{}} \toprule
experiment &$|g^{T_u}|$ $90$\% CL&   $|g^{T_d}|$ $90$\% C.L. \\
\colrule
 $^{28}$Si& $0.0065$ & $0.0065$\\
$^{76}$Ge & $0.0060$ & $0.0060$\\ \botrule
\end{tabular}\label{table:nsiN}}
\end{table}

For the case of an unparticle tensor interaction, we have found that
our constraints are more restrictive than previous analysis for values
of $d > 2.03$.

As can be seen the results are encouraging and future
$\bar \nu_e$-electron scattering experiments could give even stronger
constraints.  Another possible place to search for this type of
interaction in the future could be the coherent $\bar \nu_e$-nucleus
scattering that is also part of the TEXONO low energy neutrino physics
program ~\cite{Wong:2005vg} and other
proposals~\cite{Collar:2008zz,Anderson:2011bi,Scholberg:2005qs,Bueno:2006yq},
Table \ref{table:nsiN}.  We have shown that in this case the future
perspectives are quite encouraging since constraints to the tensorial
parameters studied in this work could be improved in more than one
order of magnitude.

\section*{Acknowledgments}
We would like to thank M. Deniz and C. Moura for useful
discussions. This work has been supported by CONACyT grant 166639 and 
SNI-Mexico.
J.B. is partially supported by UNAM-DGAPA PAPIIT IN113211. A. B.
acknowledges RED-FAE CONACYT for Postdoctoral Grant


\end{document}